\documentclass[twocolumn,floatfix,aps,prb,showpacs,superscriptaddress]{revtex4}
\usepackage{graphicx,epsf}

\usepackage{rotating}
\usepackage{dcolumn}
\usepackage{bm}
\usepackage{amsmath}
\usepackage[up]{subfigure}
\begin{document}

\newcommand{\sbar}{\bar{s}}
\newcommand{\totder}[2]{\frac{\mathrm{d}#1}{\mathrm{d}#2}}

\preprint{version of \today}

\title{The first-principles study of thermodynamical properties of
random magnetic overlayers on fcc-Cu(001) substrate }

%%\title{The first-principles study of thermodynamical properties of
%%magnetic overlayers on fcc-Cu(001) substrate: iron-cobalt overlayer
%%and partial coverage by iron atoms}

%%\title{Multiscale study of thermodynamical properties of
%%fcc-(Fe,X)/Cu(001) overlayers (X=Co,vacancy)}

%\title{Critical temperatures of random iron-cobalt overlayers
%on the fcc-Cu(001) substrate}

\author{Martin Ma\v{s}\'{\i}n
}
\affiliation{Institute of Physics, Academy of Science of the Czech Republic, Na Slovance 2, Prague, 18221, Czech Republic}

\author{Lars Bergqvist}
\affiliation{Department of Materials Science and Engineering, KTH Royal Institute of Technology,
Brinellv. 23, SE-100 44, Stockholm, Sweden}

\author{Josef Kudrnovsk\'{y} }
\affiliation{Institute of Physics, Academy of Science of the Czech Republic, Na Slovance 2, Prague, 18221, Czech Republic}

\author{Miroslav Kotrla
\footnote{Corresponding author. E-mail: {\tt kotrla@fzu.cz}}
}
\affiliation{Institute of Physics, Academy of Science of the Czech Republic, Na Slovance 2, Prague, 18221, Czech Republic}

\author{V\'{a}clav Drchal}
\affiliation{Institute of Physics, Academy of Science of the Czech Republic, Na Slovance 2, Prague, 18221, Czech Republic}

\date{\today}

\begin{abstract}
We present the theoretical study of thermodynamical properties of
fcc-Cu(001) substrate covered by iron-cobalt monolayer as well as
by incomplete iron layer. The effective two-dimensional Heisenberg
Hamiltonian is constructed from first principles and properties of
exchange interactions are investigated. The Curie temperatures are
estimated using the Monte-Carlo (MC) simulations and compared with a
simplified approach using the random-phase approximation (RPA) in
connection with the virtual-crystal approach (VCA)  to treat
randomness in exchange integrals.  Calculations indicate a weak
maximum of the Curie temperature as a function of composition of the
iron-cobalt overlayer. While a good quantitative agreement
between RPA-VCA and MC was found for iron-cobalt monolayer,
the RPA-VCA approach fails quantitatively for low coverage due
to the magnetic percolation effect.
We also present the study of the effect of alloy disorder
on the shape of magnon spectra of random overlayers.
\end{abstract}

%75.70.Ak        Magnetic properties of monolayers and thin films
%75.50.Bb        Fe and its alloys
%75.30.Et        Exchange and superexchange interactions
%75.40.-s        Critical-point effects, specific heats, short-range order
%
%05.10.Ln Monte Carlo methods
%
%

\maketitle
\section{Introduction}
\label{sec:intro}
The Curie temperature is one of the most important characteristics
of ferromagnetic materials.
Its  parameter-free determination for bulk ferromagnets and their
alloys has progressed in the last decade.
 A reasonable agreement between calculated and experimental
Curie temperatures was found for transition metal ferromagnets,
some ordered and disordered transition metal alloys (e.g., Ni-based
fcc-alloys), some f-metals (e.g., hcp-Gd, bcc-Eu), diluted magnetic
semiconductors (e.g., GaMnAs) or Heusler alloys (see, e.g.,
a recent review) \cite{tcrev}.
This progress was due to a combination of the first-principles
determination of parameters of a (classical) Heisenberg Hamiltonian
and its study using sophisticated statistical methods like, e.g.,
the random-phase approximation (RPA) or Monte-Carlo (MC)
simulations.
 It should be noted, however, that there are magnetic systems
for which the above approach has certain limitations or even fails.
The present approach assumes the existence of robust moments
and dominating effect of pair exchange interactions.
Systems with induced moments, (e.g., FeRh), magnets with more
complex than pair interactions, etc., are just few such cases,
where the above approach is not successful.
It is known that magnetic moments at system surfaces are enhanced
due to the reduced number of nearest-neighbors, which is favorable
for the validity of the Heisenberg model.

 Determination of Curie temperatures $T_{\rm C}$ of
low-dimensional systems such as ultrathin films or even random
magnetic monolayers deposited on non-magnetic substrates was
studied very rarely in the past \cite{tc2d,tcmfa} despite
their importance.
Regardless of considerable efforts in the past decade, the
parameter-free determination  of the Curie temperature of
low-dimensional systems in the framework of itinerant magnetism
remains a challenge for the theory.
There are several reasons for this: (i) Already determination of the
electronic structure of low-dimensional systems, in particular
on the first-principles level, is much more demanding as compared
to bulk systems, (ii) the presence of randomness in the system
is another complication, (iii) the statistical treatment of
the two-dimensional systems is much more delicate problem as
compared to bulk systems because in two dimensions the interactions
decay with the distance more slowly than in the bulk, and (iv) the
presence of relativistic effects has to be taken into account for
monolayers (the Mermin-Wagner theorem).

This paper is a natural extension of our previous study \cite{tc2d}
in which the Curie temperature of two-dimensional systems was
formulated for ideal, non-random systems  based on exchange
integrals determined from first-principles  (specifically for
Fe- and Co-overlayers on the fcc-Cu(001) substrate).
Here we wish to extend this study to random overlayers.
As case studies we consider (i) The random (Fe,Co)-overlayer deposited
on the fcc(001) face of Cu, and (ii) The fcc-Cu(001) substrate with
an incomplete coverage by iron atoms.
One of aims of this study is a development of a reliable scheme
for determination of $T_{\rm C}$ which employs the MC method.
This approach is accompanied by a simplified approach based on the RPA
method in which randomness in exchange integrals is treated
approximately in the virtual crystal approximation (VCA) and its
reliability and limitations are tested.
We also present the study of the effect of alloy disorder
on the shape of magnon spectra of random (Fe,Co)-overlayers obtained by
spin dynamics.

There are no experimental data for the present system to compare with
although FeCo-overlayers on various non-magnetic substrates (Cu,Pd,Rh)
were studied often with the aim to estimate the magnetic anisotropy
energies\cite{feco_over1,feco_over2,feco_over3}.
Also, there are no doubts on the importance of a parameter-free approach
to estimate $T_{\rm  C}$ of imperfect ultrathin magnetic overlayers (see,
e.g., a recent review \cite{Vaz08}).

\section{Formalism}
 \label{sec:form}
\subsection{Electronic structure and Heisenberg model}
 \label{subsec:elstruc}
The electronic structure of the system was determined in the
framework of the Green function implementation of the
scalar-relativistic tight-binding linear muffin-tin orbital
method (TB-LMTO) in which the effect of the semi-infinite substrate was
included properly in the framework of the surface Green function
(SGF) approach while the disorder in the overlayer was treated
in terms of the coherent potential approximation (CPA).
The case of partial coverage of fcc-Cu(001) by iron atoms is
simulated as a random overlayer consisting of iron atoms and
vacancies described by empty spheres.
The vacuum above the overlayer was simulated by empty spheres (ES).
Electronic relaxations were allowed in four empty spheres adjoining
the overlayer, the overlayer itself, and in five adjoining Cu substrate
layers.
This finite set of layers was sandwiched selfconsistently between a frozen
semi-infinite fcc-Co(001) and the semi-infinite vacuum including
the effect of the dipole surface barrier.
Possible small layer relaxations between overlayer and substrate
were neglected although the present approach allows to include them
\cite{feir}.
We refer the reader for more details to Refs.~\onlinecite{over1,over2}.

An important advantage of the TB-LMTO-SGF approach is a possibility to
estimate exchange interactions between magnetic atoms in the overlayer
by a straightforward generalization of the well-approved bulk concept
\cite{tcrev,lie}.
The exchange integrals $J^{Q,Q'}_{i,j}$ between sites $i,j$ occupied
by atoms $Q$ and $Q'$ ($Q,Q'$=Fe,Co or $Q,Q'$=Fe,vacancy) in the
magnetic overlayer may be expressed as follows \cite{tcrev}
\begin{equation}
J^{Q,Q'}_{i,j}  =
\frac{1}{4 \pi} \, {\rm Im}
\int_{C} \, {\rm tr}_L
\left[ {\delta^{Q}_{i}(z)} \,
{\bar g}^{\uparrow}_{i,j}(z) \,
{\delta^{ Q'}_{j}(z)} \,
{\bar g}^{\downarrow}_{j,i}(z)
\right] \, {\rm d} z \, .
\label{eqJ}
\end{equation}
Here, the trace extends over $s,p,d-$basis set, the quantities $\delta^{
Q}_{i}$ are proportional to the calculated exchange splitting, and the
(auxiliary) Green function ${\bar g}^{\sigma}_{i,j}$ describes the
propagation of electrons of a given spin
($\sigma=\uparrow,\downarrow$) between sites $i,j$ in a random
overlayer (the bar denotes the CPA configurational averaging).  The
integration path $C$ in the complex plane starts below the bottom of the
valence band and ends at the Fermi energy.  It should be noted that
both the direct propagation of electrons in the random magnetic
overlayer and the indirect one in the semi-infinite Cu-substrate are
included in Eq.~(\ref{eqJ}) on an equal footing. Finally, the
disorder-induced vertex corrections due to the correlated motion of two
electrons in a random overlayer can be neglected in a reasonable
approximation due to the vertex-cancelation theorem \cite{vct}.

Once the exchange interactions were known, we constructed a
two-dimensional (2D) random classical Heisenberg Hamiltonian to
describe the magnetic behavior of the random (Fe,$X$)-overlayer
($X$=Co,vacancy) on a non-magnetic fcc-Cu(001) substrate
\begin{eqnarray}
H = -\ \sum_{Q,Q'={\rm Fe,Co}} \sum_{i \neq j} \eta^{Q}_{i} \,
J^{Q,Q'}_{ij} \eta^{Q'}_{j} {\bf e}_{i} \cdot {\bf e}_{j} \,\nonumber\\
+  \sum_{Q = {\rm Fe, Co}}     \Delta^Q \sum_i (\eta^{Q}_{i})^2 ({\bf e}_i^z)^2 \, .
\label{eqH}
\end{eqnarray}

In Eq.~(\ref{eqH}), ${\bf e}_i$ denotes a unit vector with the
direction of the local magnetic moment at the site $i$ and
$\eta^{Q}_{i}$ is the occupation index which equals 1 if the
site $i$ is occupied by the atom $Q$ and zero otherwise.  The second term
is an uniaxial anisotropy of strength $\Delta^Q$ with an easy axis out of plane.
By construction, the value of the corresponding magnetic moment
is included in the definition of $J^{Q,Q'}_{ij}$, and  positive
(negative) values denote  FM (AFM) couplings.
Because exchange integrals for two-dimensional case decay with the
distance $d$ between sites $i$ and $j$ more slowly as compared to
the bulk, a large number of shells is needed to obtain well
converged results, in particular for non-random cases of pure
Fe- and Co-overlayers.
Up to 90 shells were included in calculations of $T_{\rm C}$ in the
framework of the RPA and MC methods.
The effect of small induced moments on substrate atoms was neglected
(moments are of order few hundredths of $\mu_{\rm B}$).

\subsection{Statistical treatment}
 \label{subsec:stat}
 \subsubsection{Random Phase Approximation}
 \label{subsubsec:rpa}

The expression for $T_{\rm C}$ in the RPA is a generalization of
its bulk counterpart \cite{tcrev} to the case of random magnetic
overlayers \cite{tc2d}: (i) A vanishing $T_{\rm C}$ is obtained in
agreement with the Mermin-Wagner theorem for vanishing anisotropy
energy $\Delta^{ Q}$;
(ii) The anisotropy energy is taken here as an adjustable parameter
which we have identified with the dipolar energy and used the same
values as in Ref.~\onlinecite{tc2d}.
This is not a serious problem as $T_{\rm C}$ has only a weak logarithmic
dependence on $\Delta^{ Q}$ , cf. Ref. \onlinecite{tc2d}; and (iii) We have
averaged three exchange integrals $J^{ Q,Q'}_{ij}$
($Q,Q'$=Fe,Co or $Q,Q'$=Fe,vacancy) and introduced the effective
non-random exchange integrals $J^{\rm eff}_{ij}$ defined as
\begin{equation}\label{jeff}
J^{\rm eff}_{ij} =
x^{2} J^{\rm Fe,Fe}_{ij} + x(1-x) (J^{{\rm Fe},X}_{ij} +
J^{X,{\rm Fe}}_{ij}) + (1-x)^{2} J^{X,X}_{ij}
\end{equation}
which depend on the actual composition of random alloys
Fe$_{x}X_{1-x}$ ($X$=Co,vacancy).
This is the virtual-crystal approximation (VCA) and we have tested its
applicability by performing MC simulations in which this approximation
is not used (cf. Sec. \ref{subsec:resTcFeCo}).
Using the VCA, the Curie temperature is
\begin{equation}\label{tcrpa}
(k_{B} \, T_{\rm C}^{\rm RPA})^{-1} = \frac{3}{2} \frac{1}{N_{\|}} \;
\sum_{\bf q_{\|}} \, [ \Delta +  J^{\rm eff}({\bf 0}) -
J^{\rm eff}({\bf q_{\|}}) ]^{-1} \, ,
\end{equation}
where $J^{\rm eff}({\bf q_{\|}})$ is the lattice Fourier transform
of the effective exchange integrals $J^{\rm eff}_{ij}$.

\subsubsection{Monte Carlo simulation}
 \label{subsec:MC}

In present work, we use the UppASD package \cite{UppASD},
developed at the Uppsala University for the study of magnetic materials.
This package also contains MC code with the Metropolis algorithm
which we utilized. The sampling in the Metropolis algorithm is controlled
by the transition rate $w(i,f)$  from an initial state $i$ to a final state $f$
which depends on energies  $E_i$ and $E_f$ of an initial and a final
state, respectively. The energies  $E_i$ and $E_f$ were calculated
using the Heisenberg Hamiltonian (\ref{eqH}).\\

The number of trial moves in every Monte Carlo step (MCS)
corresponds to the number of spins in the overlayer. The substrate is
treated as rigid, adatoms do not diffuse, but rotation of spins of all
adatoms is allowed. We start the simulation with a system of randomly
distributed spin vectors in the overlayer. In the first phase of simulation,
we only equilibrate the spin system; we use typically 10,000 MCS. After
that, we perform the measurement of observable quantities; we
measure over 50,000 MCS. We average calculated quantities over
several independent runs. In the case of a random alloy, we employ
several (typically 20) different random distributions of Fe and Co
adatoms  or Fe adatoms and vacancies in the overlayer.

\begin{figure}
\center \includegraphics[width=8cm]{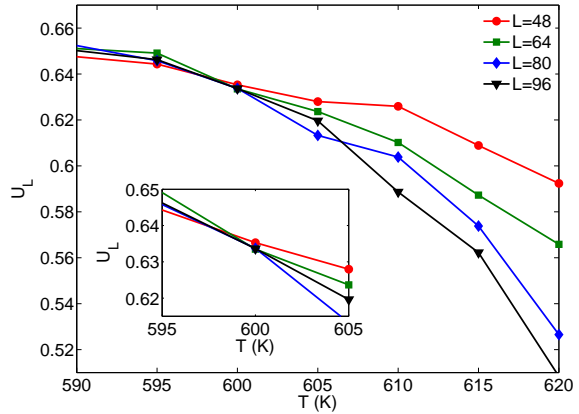}
\caption{(Color online) Temperature dependence of Binder cumulants $U_L$ for different system sizes
 in the case of fcc (Fe,Co)/Cu(001)
  random overlayer with the concentration $x=0.5$.  }
\label{fig_UL}
\end{figure}

One can use several methods for an estimation of the Curie
temperature. We illustrate these possibilities in the case of a random
(Fe,Co)-overlayer with the concentration $x = 0.5$. One option is to
employ the fourth--order size--dependent Binder cumulant
\cite{binder81} $U_{L} = 1- \left< m_z^{4} \right> / \left[ 3 \left< m_z ^ 2
\right>^2 \right]$. Here, $m_z$ is  the z-component of the
magnetization and $L$ is the system size. This approach turned out to
be quite useful in the past, though some limitations exist \cite{holt}. The
shape of temperature dependence of the magnetization is influenced by
the size $L$. It leads often to the situation that cumulants for different
system sizes cross in one point corresponding to the Curie temperature
\cite{binder81}. We calculated cumulants in our case, see
Fig.~\ref{fig_UL}.
Unfortunately, we have found that the identification of a crossing
point is not so clear as in the case of three dimensional systems.

Other methods are based on finding of a singularity point in the susceptibility $\chi$
or in the specific heat $C$. In the simulations, one can directly obtain only
size-dependent quantities $\chi_L$ or $C_L$. One locates the temperature $T_\chi^L$ or $T_C^L$
of a local maximum of the susceptibility or specific heat for the size $L$ and then
performs extrapolation using corresponding scaling relation.

In this paper, we used mainly susceptibility (see Secs. \ref{subsec:resTcFeCo} and \ref{subsec:resTcFeVacance}).
After location a temperature $T_{\chi}^L$ of a local maximum of the susceptibility for the size $L$, we utilized the scaling relation
\begin{equation}
T_{\chi}^L  \approx T_{\chi}  + \lambda L^{ -\frac{1}{\nu} },
\end{equation}
where $T_{\chi}$ is an estimate of Curie temperature for the infinite system.
More specifically, the critical exponent $\nu = 1$ for the two dimensional Heisenberg model
with uniaxial anisotropy is known \cite{2dHeis}.
The system falls into the same universality class as the
two-dimensional Ising model, for which all the critical
exponents have analytically known values.

\begin{figure}
\center \includegraphics[width=8cm]{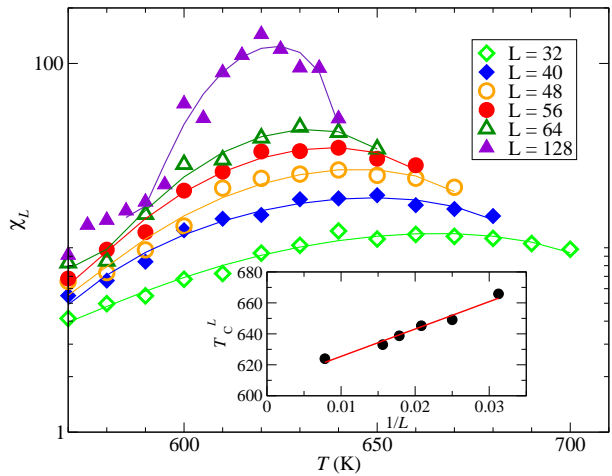}
\caption{(Color online) Temperature dependence of size-dependent susceptibilities $\chi_L$ for fcc (Fe,Co)/Cu(001)
  random overlayer with the concentration $x=0.5$.  }
\label{fig_chi}
\end{figure}

For the comparison with the method of cumulants, we evaluated $T_{\rm C}$ by the calculation of the size-dependent susceptibilities $\chi_L$ for fcc (Fe,Co)/Cu(001)
  random overlayer with the concentration $x=0.5$ in Fig.~\ref{fig_chi}.
We carried out simulations for several system sizes (typically ranging from $L=16$ to
$L=128$ and then we estimated $T_{\rm C}$ as an extrapolation of $T_C^L$
using linear regression for infinite $L$.

Similarly we calculate the size-dependent specific heat
(figure is not presented).

\subsection{Atomistic spin dynamics}
 \label{subsec:spindyn}

Using the generalized Hamiltonian $H$ ,
Eq.~(\ref{eqH}), as a starting point, the temporal evolution of the atomic moments,
$\mathbf {m}_i$, where $\mathbf {m}_i=|m_i|{\bf e}_i$ and $|m_i|$ is the amplitude of the magnetic moment,
at finite temperature is governed by Langevin dynamics through coupled stochastic differential equations of the Landau-Lifshitz form,

\begin{eqnarray}
\!\!\! \frac{\partial\mathbf{m}_i}{\partial t} = &&- \frac{\gamma}{(1+\alpha ^2)}  \mathbf{m}_i \times \left[\mathbf{B}_i + \mathbf{b}_i(t)\right]\nonumber\\
&& - \gamma\frac{\alpha}{|m| (1+\alpha ^2)} \mathbf{m}_i \times \left\{ \mathbf{m}_i \times\left[\mathbf{B}_i + \mathbf{b}_i(t)\right] \right\}, \quad
\label{eqn:LLG}
\end{eqnarray}

\noindent where $\gamma$ is the electron gyromagnetic ratio,
and  $\mathbf{B}_i$ is the effective field produced by all other moments.
Effects of temperature are included via adding a stochastic Gaussian--distributed
magnetic field $\mathbf{b}_i(t)$, the strength is controlled by the Gilbert
damping parameter $\alpha$. In principle, the Gilbert damping
parameter could be calculated by means of electronic structure
methods, but here it is considered as a parameter. More information
about the numerical details and integrating scheme for solving
Eq.(\ref{eqn:LLG}) can be found in Ref.~\onlinecite{Mentink2010}.

The principal advantage of combining first-principles calculations with the atomistic spin
dynamics (ASD) approach is that it allows to address the dynamical properties of spin
systems at finite temperatures~\cite{Skubic2008,Tao2005,Chen1994}.
Two important quantities we focus on are the space- and time-displaced correlation function,

\begin{equation}
  C^\nu_{i j} (t) = \langle {\bf m}_{i}^{\nu}(t) {\bf m}_{j}^{\nu}(0) \rangle
  - \langle {\bf m}_{i}^{\nu}(t) \rangle \langle {\bf m}_{j}^{\nu}(0) \rangle,
%  C^\nu (\mathbf{r}-\mathbf{r'},t) = \langle {\bf m}_{\mathbf{r}}^k(t) {\bf m}_{\mathbf{r'}}^k(0) \rangle - \langle {\bf m}_{\mathbf{r}}^k(t) \rangle \langle {\bf m}_{\mathbf{r'}}^k(0) \rangle,
\end{equation}

\noindent where the angular brackets denote an ensemble average and $\nu$ is the cartesian component.
%, and ${\bf r}$ is a position of a magnetic moment vector ${\bf m}_i$ in the lattice.
Its Fourier transform is the dynamical structure factor

\begin{equation}\label{structure_factor}
  S^{\nu}(\mathbf{q},\omega) = \frac{1}{\sqrt{2\pi}N} \sum_{i j} e^{i\mathbf{q}\cdot(\mathbf{R_i}-\mathbf{R_j})}
  \int_{-\infty}^{\infty} e^{i\omega t} C^{\nu}_{i j} (t) dt,
\end{equation}

\noindent where $\mathbf{q}$ and $\omega$ are the momentum and energy transfer, respectively.
$S(\mathbf{q},\omega)$ is the quantity probed in neutron scattering experiments of bulk systems,
and can analogously be applied to spin-polarized electron energy loss spectroscopy (SPEELS) measurements.
By plotting the peak positions of the structure factor along particular directions in reciprocal
space, the magnon dispersions may be obtained ~\cite{Skubic2008,Tao2005,Chen1994}.

\section{Results}

\subsection{Exchange interactions}
 \label{subsec:exch}

\begin{figure}
\center \includegraphics[width=8cm]{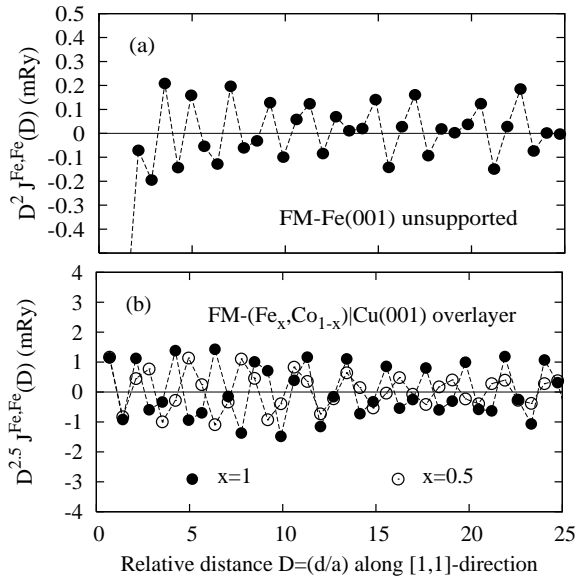}
\caption{The exchange integrals as a function of the distance along
 the direction [11] in the magnetic layer: (a) The unsupported
 Fe-layer, and (ii) The case of Fe- and Fe$_{0.5}$Co$_{0.5}$ overlayers
 on the fcc Cu(001) substrate. The corresponding exchange integrals
 are multiplied by a suitable power of distance to highlight their decay
 exponent.  Note that the first two exchange integrals in (a) are out
 of frame.        }
\label{fig_dist}
\end{figure}

In this section we wish to illustrate some general features of
exchange interactions for 2D systems.
We shall start with the study of the dependence of exchange integrals
$J^{\rm Fe, Fe}_{ij}$ on the distance between Fe atoms in the fcc(001)
monolayer along the direction [1,1] in various cases, namely (i) The
isolated (unsupported) Fe-layer, (ii) The Fe-overlayer on fcc-Cu(001),
 and (iii) The random (Fe$_{0.5}$,Co$_{0.5}$)-overlayer on fcc-Cu(001).
In this way we can study the effect of indirect interactions of
Fe-atoms via the substrate (missing for unsupported layer) as well
as the effect of disorder.
The case of unsupported Fe-layer was studied using the same model in
which, however, the iron overlayer was separated from the substrate
by eight layers of empty spheres from the fcc-Cu(001) substrate.
It is well-known that the exchange integrals in bulk ferromagnets
decay with distance $d$ as $d^{-3}$.
It is seen from Fig.~\ref{fig_dist}a that for the unsupported layer
the decay with distance is much slower, namely, proportional to $d^{-2}$.

The effect of the substrate is striking as illustrated in
Fig.~\ref{fig_dist}b for Fe-overlayer on fcc-Cu(001).
The decay of exchange interactions with distance $d$ is approximately
$d^{-2.5}$ as a result of interactions via the substrate.
Also illustrated in Fig.~\ref{fig_dist}b is the additional exponential
decay of exchange interactions due to the alloy disorder for the case
of (Fe$_{0.5}$Co$_{0.5}$) overlayer on the substrate.

\begin{figure}
\center \includegraphics[width=8cm]{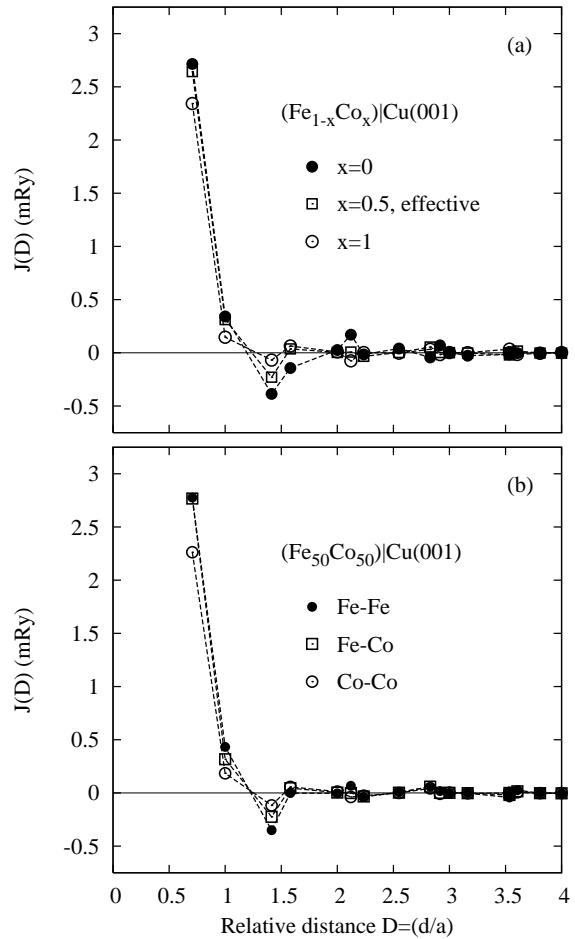}
\caption{The exchange integrals in (Fe$_{x}$,Co$_{1-x}$) overlayer
 on the fcc-Cu(001) substrate as a function of the shell distance:
 (a) The case of pure Fe and Co overlayers and the effective VCA
 exchange integrals {\bf (\ref{jeff}) } (Fe$_{0.5}$,Co$_{0.5}$)
 overlayer (empty squares); and (b) The species-resolved exchange
 integrals $J^{ Q,Q'}_{ij}$, $Q,Q'$=Fe,Co for (Fe$_{0.5}$,Co$_{0.5}$)
 overlayer. }
\label{fig_feco}
\end{figure}

The exchange integrals for pure Fe and Co overlayers on
fcc-Cu(001) substrate are shown in  Fig.~\ref{fig_feco}a.
We observe generally larger values of exchange integrals for the
iron overlayer as compared to a cobalt one indicating a higher
Curie temperature in the former case.
Also shown are configurationally averaged integrals
$J^{ Q,Q'}_{ij}$, $Q,Q'$=Fe,Co for (Fe$_{0.5}$,Co$_{0.5}$)
overlayer (for VCA values, see Eq.~(\ref{jeff})) used in an
approximate treatment of the Curie temperature (see below).
It is interesting to note that these effective integrals are
similar to the species resolved $J^{\rm Fe,Co}_{ij}$ integrals
shown in Fig.~\ref{fig_feco}b.

Species-resolved exchange integrals for equiconcetration coverage of
fcc-Cu(001) substrate by iron and cobalt atoms are shown in
Fig.~\ref{fig_feco}b.  The remarkable feature is a relative similarity
of all three kinds of exchange integrals which is an indication of
the validity of a simplified VCA treatment as discussed below.

\begin{figure}
\center \includegraphics[width=8cm]{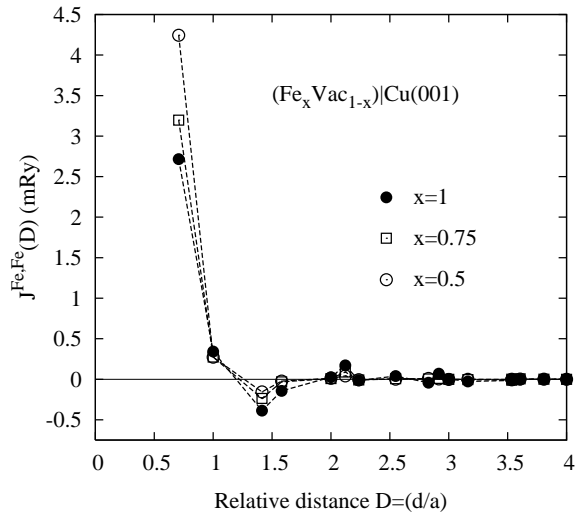}
\caption{The Fe-Fe exchange integrals as a function of the shell
 distance for three different coverages of fcc-Cu(001) surface by
 Fe atoms, indicated in the plot. }
\label{fig_fevac}
\end{figure}

Finally, in Fig.~\ref{fig_fevac} we show the dependence of exchange
integrals $J^{\rm Fe, Fe}_{ij}$ on the coverage $x$ of the fcc-Cu(001)
surface by iron atoms.
The exchange integrals increase with decreasing coverage.
The reason is their impurity character for the case of partial
coverage, namely, decreasing number of nearest neighbors which also
leads to enhanced Fe-moments as compared to a full Fe overlayer.
Mathematically, the exchange integrals are scaled by the same factor
which also increases the impurity density of states (or local moments)
as compared to the host one.
On the other hand, their effective number is proportional to $x^{2}$.
As a result, the effective interactions (see Eq.~(\ref{jeff}))
decrease with decreasing coverage indicating the monotonic decrease
of the corresponding Curie temperature.

\subsection{Curie temperature: iron-cobalt overlayer}
\label{subsec:resTcFeCo}
We have first investigated the magnetic properties of random
Fe$_{1-x}$Co$_x$ overlayer on Cu substrate using two methods:
the RPA and the MC simulation. In MC simulation we calculated the
size-dependent susceptibility and then performed extrapolation
as described in the subsection \ref{subsec:MC}.
As discussed above, in the RPA we used an approximate VCA while
MC simulations were done both for the VCA as well as for the
realistic case with three different exchange integrals
$J^{ Q,Q'}_{ij}$ ($Q,Q'$=Fe,Co).
The RPA-VCA approximation was used successfully for the calculation
of $T_{\rm C}$ of bulk Ni-rich transition metal alloys. \cite{bvca}
In Ref.~\cite{mcvca}, it was demonstrated numerically on a
simple model that VCA is a good approximation above the percolation
limit and for extended exchange integrals.
Another reason for validity of the RPA-VCA in this case is
a similarity of exchange integrals $J^{ Q,Q'}_{ij}$ ($Q,Q'$=Fe,Co)
as shown above.
On the contrary, for localized exchange integrals and a very
low concentration of magnetic impurities the VCA fails, like,
e.g., in (Ga,Mn)As and (Ga,Mn)N diluted semiconductor alloys.
\cite{mcvca}
In the present case one should expect that the VCA will be a good
approximation because of extended character of exchange integrals
$J^{ Q,Q'}_{ij}$ and their similarity for various atom
types as indicated above.

\begin{figure}
\center \includegraphics[width=8cm]{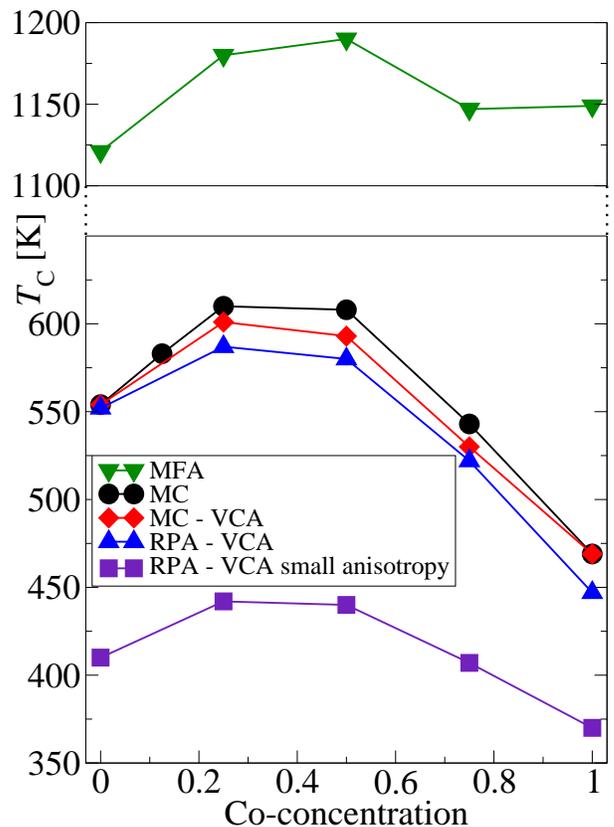}
\caption{ (Color online) Concentration dependence of $T_{\rm C}$ for fcc (Fe,Co)/Cu(001)
  random overlayer as a function of Co-concentration. Lower panel: We compare
  the MC simulations using random exchange integrals (circles)
  with the simplified MC-VCA (diamonds) and RPA-VCA (triangles and squares). In
  the VCA cases, we employ non-random alloy with effective
  concentration-dependent exchange integrals (see the text). Also shown is the RPA-VCA
  result for reduced anisotropy  (see the text).
   Upper panel shows mean field approximation.}
\label{fig_tc}
\end{figure}

The calculation confirms this prediction as it is obvious from
Fig.~\ref{fig_tc} in which we compare $T_{\rm C}$ of
fcc-(Fe,Co)/Cu(001) random overlayer over the whole concentration
range.
In the limit of Fe- and Co-overlayer on fcc-Cu(001) the present results
agree well with a previous study \cite{tc2d} and small differences are
due to the differences in technical details.
We have obtained  well-pronounced maximum in the concentration
dependence of $T_{\rm C}$. which reminds a similar maximum in the
concentration trend of random bulk  FeCo alloys in the concentration
range in which the bcc-phase exists \cite{lezaic}.
The calculated $T_{\rm C}$ depends on the values of the anisotropy
energy $\Delta$ of constituent atoms, although such dependence is
quite weak as demonstrated in Ref.~\onlinecite{tc2d}.
 We remind that this term is responsible for the formation of a
narrow gap in the spin-wave spectra which is an origin of the finite
Curie temperature for the monolayer case\cite{tc2d}.
Due to this term the Curie temperature determined in the
framework of the RPA or Monte Carlo is much lower than its mean-field
counterpart as it is obvious from Fig.~\ref{fig_tc} (see also
Ref.~\onlinecite{tcmfa}).
A reliable choice of this term is a problem.
In the present case we have chosen the same values as in
Ref.~\onlinecite{tc2d} ($\Delta^{\rm Fe}$=0.140~mRy $\Delta^{\rm Co}$=0.052~mRy)
and their average in the VCA models
 in order to have a direct comparison with the case of ideal
overlayers.
A recent study \cite{msa} indicates that this value may be
significantly smaller.
Because of this uncertainty, we have tested the robustness
of the present result with respect to various values of this term
using the RPA-VCA method.
In particular, we show in Fig.~\ref{fig_tc} results of the RPA-VCA
for the case in which we used the same values of $\Delta$ for both
Fe and Co, but reduced by an order of magnitude as compared to that for
Fe-atoms ($\Delta$=0.015~mRy).
Calculated Curie temperatures are lowered in agreement with the
previous study\cite{tc2d}, but the concentration maximum, although
less pronounced due to the same values of $\Delta$, is still present.
It should be noted that a reliable experimental estimate of the
Curie temperature of monolayers is still a challenge.

 We note that  similar maxima in the concentration trend of
$T_{\rm C}$ exist in random bulk bcc-FeCo \cite{lezaic}
and fcc-NiFe alloys \cite{bvca}.
In the former case the total moment has a weak maximum for
about 30\% of Co \cite{feco_2d3d}, in the latter case the total
moment varies almost linearly with the content \cite{bvca}
similarly like in the present overlayer case. \cite{feco_2d3d}
This means that there is no simple relation between concentration
trends of total moments and Curie temperatures.
The Curie temperature is determined by the exchange
integrals.
It should be noted that a discussed concentration maximum of
$T_{\rm C}$ is seen already in their mean-field values,
Fig.~\ref{fig_tc}, which, in turn, are directly proportional
to the sum of averaged exchange integrals, Eq.~(\ref{jeff}).
This can be considered as a precursor of the concentration
maximum of $T_{\rm C}$.

We can conclude that a critical comparison of calculated Curie
temperatures of random magnetic overlayers on nonmagnetic
substrates can be conveniently performed using a simplified RPA-VCA
approach which is significantly more numerically efficient than
MC simulations.
This is mostly due to the fact that exchange integrals for various
species are rather similar.
Using the fast RPA-VCA approach we have verified that due to
a slow decay of exchange integrals with distance one should
carefully check the convergence of results with the number
of shells included in simulations.

\subsection{Curie temperature: incomplete coverage by iron}
\label{subsec:resTcFeVacance} The next case is the model of partial
coverage of the fcc-Cu(001) substrate by iron atoms. In this case,
contrary to the previous model of (Fe,Co)-overlayer, the only non-zero
exchange integrals are those among Fe atoms.
The effective exchange
integral (VCA) is thus $J^{\rm eff}_{ij}= x^{2} J^{\rm Fe,Fe}_{ij}$, where
$x$ denotes the coverage ($x$=1 corresponds to ideal
Fe-overlayer). It should be noted that for 2D-systems the effect of
percolation may lead to reduction of Curie temperature to zero value. In
particular, it is known that the site percolation threshold for the square
lattice with interaction up to next nearest neighbors is $p_c \approx$
0.592 and that the percolation threshold may progressively decrease
with the range of interaction \cite{percol} provided the strength of
interaction is independent of a range of neighbor bonds. Despite the
fact that exchange integrals are relatively long-ranged, one thus should
expect that the VCA will represent a substantial approximation, at least at
low coverages.

We have estimated the Curie temperature of fcc-Fe$_{x}$/Cu(001)
system for coverages $x$=1, 0.75, 0.5, and 0.25.
The results are summarized  in the Table~1.
Both the RPA-VCA and MC simulations give qualitatively the same
result, namely, an almost linear decrease of the Curie temperature
with decreasing coverage.
On the other hand, the RPA-VCA overestimates the Curie temperature
with decreasing coverage as expected and for very low coverage
($x$=0.25) it gives even the qualitatively wrong result as the
MC leads to a collapse of the long-range magnetic order.
One can thus conclude that for systems with very different exchange
integrals one should be careful when using simplified approaches
like the RPA-VCA.
Another example of such system can be, e.g., the surface alloy
fcc-(Cu,Mn)/Cu(001) with negligible exchange integrals between Cu
atoms and also between Cu and Mn atoms due to a negligible magnetization
of Cu atoms,
On the other hand, the 2D-generalization of the random RPA approach
\cite{rrpa} which was successfully applied to the dilute magnetic
semiconductors can be a fast and reliable counterpart to the
numerically demanding MC simulations which, on the other hand, are
the most reliable.

%\begin{table*}
\begin{table}
\caption{\label{tab:table1}Curie temperatures of overlayer with incomplete
coverage by iron.  Values of $T_{\rm C}$ obtained by MC simulation and RPA calculation
for different concentrations $x$ are shown.}
	\begin{ruledtabular}
	\centering
		 \begin{tabular}{p{1.5in}p{1.0in}p{1.3in}p{0.6in}}
$x$  &	$T_{\rm C}^{\rm MC}$ (K) & $T_{\rm C}^{\rm RPA}$ 	(K)  \\
\hline
1.0 &  552   &	552\\
0.75 & 361    & 415\\
0.5 & 140    &	269\\
0.25 & --    & 131\\
 		\end{tabular}
	\end{ruledtabular}	
\end{table}
%\end{table*}

\subsection{Magnon spectra}
\label{sec:magnon}
%50+50 alloy, effect of temperature 5 K and 300 K, different damping

\begin{figure}
\center \includegraphics[scale=0.35]{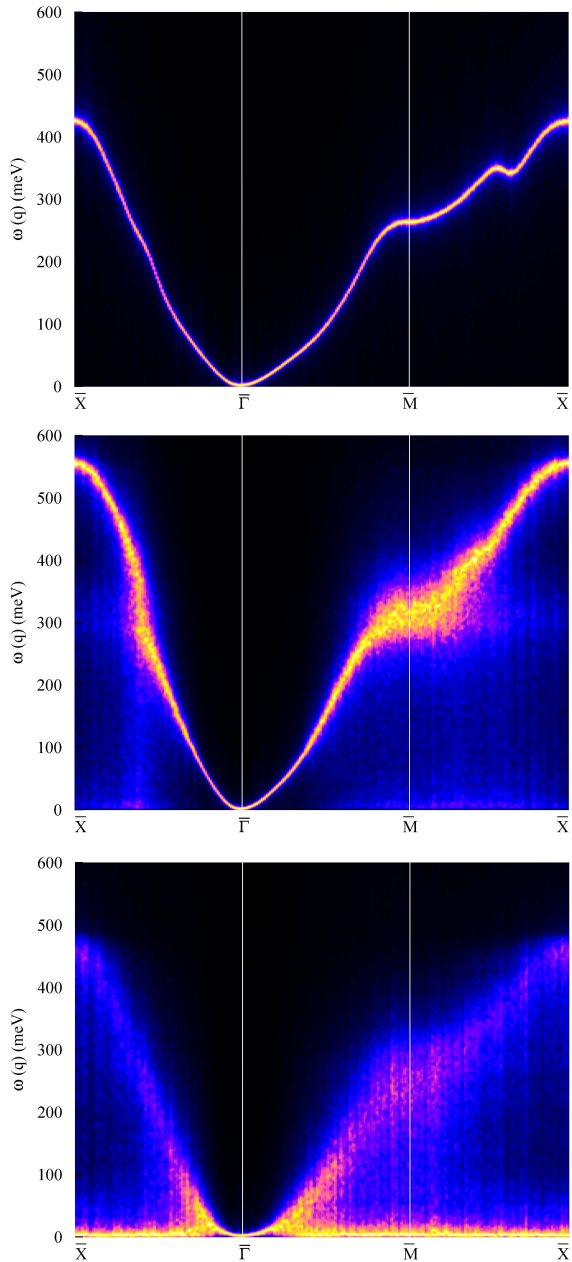}
\caption{(Color online) Magnon spectra for Fe/Cu(001) (upper panel),
(Fe$_{0.50}$,Co$_{0.50}$)/Cu(001) (middle panel) and (Fe$_{0.75}$,Vac$_{0.25}$)/Cu(001) (lower panel)
evaluated at $T=5$ K, and for $\alpha$=0.005.  }
\label{fig_ma1}
\end{figure}

Progress in experimental techniques in the recent years has made it possible
to measure magnon dispersion even in ultrathin magnetic overlayers, such as
a single Fe monolayer on W(110), using spin-polarized electron energy loss
spectroscopy (SPEELS) measurements \cite{Prokop2009}.
Although the present random iron-cobalt overlayer has not yet been measured,
our calculations hopefully could motivate such a study.
In principle, there are at least
three sources of broadening of the magnon dispersion, namely, the alloy disorder,
the transversal temperature
fluctuations of the magnetic moments originating from the coupled thermal bath,
and the longitudinal Stoner excitations. At present, the ASD formalism does not include
Stoner excitations, but for low temperatures and smaller wavevectors, the density of
magnons dominates over Stoner excitations. Here, we focus on the broadening originating
from alloy disorder by calculating magnon dispersions for the following cases (i) pure Fe,
(ii) Fe$_{50}$Co$_{50}$, and (iii) Fe$_{75}$Vac$_{25}$ overlayer on top of Cu(001),
 as displayed in Fig.\ref{fig_ma1}. The color is a measure of magnon damping (the full width at half maximum
of the magnon spectral function).
  In all three cases, the temperature is fixed at T=5K and $\alpha$=0.005. We probe magnons
  in the fcc(001) two-dimensional Brillouin zone by following
the path $\bar X - \bar \Gamma - \bar M - \bar X$. The $\bar X$ and $\bar M$ points
correspond to p(2x1) and c(2x2) antiferromagnetic structures, respectively.

The magnon spectra for the Fe overlayer show very little broadening throughout
the Brillouin zone as expected, since there is no alloy disorder in this case.
Alloying Fe with Co changes the situation. Here the spectra are more diffuse due
to the alloy disorder. It is worth noting that the broadening is most pronounced
around the $\bar M$ point in the Brillouin zone.
The higher frequencies at the $\bar X$ and $\bar M$ points
are consistent with the higher $T_{\rm C}$ found in Fe$_{0.5}$Co$_{0.5}$ overlayer as compared
with the pure Fe case. The difference
 between magnetic properties of
 Fe and Co atoms is not very big and there
are still distinct magnon excitations visible in the whole BZ. However, an extreme situation
is encountered in the dilute case (Fe$_{75}$Vac$_{25}$ alloy).
Here the alloy disorder is large causing a very diffuse magnon spectra. A similar result has been found in three--dimensional diluted magnets by random phase approximation\cite{Bouzerar}.
One can say that magnon excitations in this case are strongly damped and existing
only for small wave vectors, i.e., for longwave magnons which are less
sensitive to local site disorder.
In this  incomplete  Fe coverage ($x$=0.75), we observe a dispersionless
 excitation with the energy close to zero. It corresponds to
 localized magnons that appear close and below the percolation
 threshold. This phenomenon was discussed by Chakraborty and Bouzerar [\onlinecite{Bouzerar}] in
 a three--dimensional case, but the same arguments are valid also
 in two dimensions. We note that the percolation concentrations for
 fcc-lattice and for the square lattice are around 0.2 and 0.5,
 respectively.

\section{Discussion and conclusion}

We have presented the first-principles theory of thermodynamical properties
of random magnetic overlayers on non-magnetic metallic substrates and
applied it to two systems, namely, to a random iron-cobalt monolayer on
the fcc-Cu(001) substrate and to the case of partial coverage of
fcc-(001) face of copper by iron atoms.
Atomistic spin dynamics simulations were used to predict magnon
dispersions in random overlayers, where broadening from
alloy disorder was quantified.

The main conclusions from our study are:
(i) The exchange integrals of magnetic overlayers decay
 more slowly as compared to the bulk cases and large number of shells has to
 be included in the statistical treatment to obtain well converged result;
(ii) Contrary to the bulk case, we have found that the most efficient and
reliable approach to estimate the Curie temperature of magnetic overlayers
is the estimate of the susceptibility combined with the scaling relation rather
than the cumulant method;
(iii) The simplified, but numerically efficient RPA-VCA approach was tested
and limits of its applicability were established. It was found that
 the RPA-VCA is applicable for systems with similar exchange integrals
(like, e.g., FeCo-overlayer), but it fails for systems with very different
values of exchange integrals (like, e.g., incomplete Fe-overlayer);
(iv) Calculations indicate the presence of a maximum in the concentration
dependence of the Curie temperature similar to that observed in bulk
bcc-FeCo alloys.

\section{Acknowledgements}

The authors acknowledge the support from
%the Grant Agency of the Czech Republic
Czech Science Foundation
under Contract No. 202/09/0775.
L.B. acknowledges support
from Swedish e-Science Research Centre (SeRC), Swedish Research Council (VR)
and G\"oran Gustafssons Foundation.

%LATER USE BIBTEX
%\bibliography{Fe-Co}
\bibliography{FeCo}

\end{document}